 \documentclass[aps, pra, 10pt,twocolumn, showpacs]{revtex4-1}

\usepackage{epsfig,graphics}

\newcommand{\njp}{New J. Phys.~}

\newcommand{\jpb}{J. Phys. B~}
\newcommand{\pla}{Phys. Lett. A~}

\def \etal{{\em et al.}}

\def \azh{\hat{a}_{{\rm 0H}}^\dagger}
\def \azv{\hat{a}_{{\rm 0V}}^\dagger}
\def \aah{\hat{a}_{{\rm 1H}}^\dagger}
\def \aav{\hat{a}_{{\rm 1V}}^\dagger}
\def \abh{\hat{a}_{{\rm 2H}}^\dagger}
\def \abv{\hat{a}_{{\rm 2V}}^\dagger}
\def \abph{\hat{a}_{{\rm 2^{\prime}H}}^\dagger}

\def \ach{\hat{a}_{{\rm 1^{\prime}H}}^\dagger}
\def \acv{\hat{a}_{{\rm 1^{\prime}V}}^\dagger}
\def \adh{\hat{a}_{{\rm 2^{\prime}H}}^\dagger}
\def \adv{\hat{a}_{{\rm 2^{\prime}V}}^\dagger}

\def \tr{{\rm Tr}\,}
\def \>{\rangle}
\def \<{\langle}
\def \E{{\rm e}}

\newcommand\ket[1]{\ensuremath{|#1\rangle}}
\newcommand\bra[1]{\ensuremath{\langle#1|}}

\begin{document}

\title{Cloning of arbitrary mirror-symmetric distributions on Bloch
sphere: Optimality proof and proposal for practical photonic
realization}

\author{Karol Bartkiewicz}

\affiliation{Faculty of Physics, Adam Mickiewicz University,
61-614 Pozna\'n, Poland}

\author{Adam Miranowicz}

\affiliation{Faculty of Physics, Adam Mickiewicz University,
61-614 Pozna\'n, Poland}

\begin{abstract}

We study state-dependent quantum cloning which can outperform
universal cloning. This is possible by using some a priori
information on a given quantum state to be cloned. Specifically,
we propose a generalization and optical implementation of quantum
optimal mirror phase-covariant cloning, which refers to optimal
cloning of sets of qubits of known modulus of expectation value of
Pauli's $Z$ operator. Our results can be applied for cloning of an
arbitrary mirror-symmetric distribution of qubits on Bloch sphere
including in special cases the universal cloning and
phase-covariant cloning. We show that the cloning is optimal by
adapting our former optimality proof for axisymmetric cloning
[Phys. Rev. 82, 042330 (2010)]. Moreover, we propose an optical
realization of the optimal mirror phase-covariant $1\to2$ cloning
of a qubit, for which the mean probability of successful cloning
varies from $1/6$ to $1/3$ depending on prior information on the
set of qubits to be cloned. The qubits are represented by
polarization states of photons generated by the type-I spontaneous
parametric down-conversion. The scheme is based on the
interference of two photons on an unbalanced
polarization-dependent beam splitter with different splitting
ratios for vertical and horizontal polarization components and the
additional application of feedforward by means of Pockels cells.
The experimental feasibility of the proposed setup is carefully
studied including various kinds of imperfections and losses
including: (i) finite efficiency of generating a pair of entangled
photons in the type-I spontaneous parametric down conversion, (ii)
the influence of choosing various splitting ratios of the
unbalanced beam splitter, (iii) the application of
conventional and single-photon discriminating detectors, (iv) dark
counts and finite efficiency of the detectors.

\end{abstract}

\date{\today}
\pagestyle{plain} \pagenumbering{arabic}

 \maketitle

\section{Introduction}

The no-cloning theorem~\cite{Zurek,Dieks} tells that unknown
quantum states cannot be copied perfectly, which is implied by the
linearity of quantum mechanics. The no-cloning theorem guaranties,
e.g., the security (or privacy) of quantum communication protocols
including quantum key distribution and excludes naive protocols of
superluminal communication with entangled particles.

As perfect quantum cloning is impossible, much attention has been
devoted to approximate~\cite{Dieks,Buzek96} and
probabilistic~\cite{Duan98} quantum cloning. Such studies are
especially important for quantum cryptography~\cite{Fuchs97}, but
also for quantum communication~\cite{Bruss01} and
computation~\cite{Galvao00}.

It is worth noting that quantum cloning is not only of theoretical
interest. In fact, a few experimental realizations of quantum
cloning have been reported~\cite{Experiment}. In particular,
quantum cloning with prior partial information, which is the main
topic of our paper, was experimentally demonstrated using nuclear
magnetic resonance~\cite{ExperimentPCC2} and optical
systems~\cite{Cernoch2006,ExperimentPCC}. Also quantum-dot
implementations of cloning machines were
considered~\cite{ExperimentQD}.

The first $1\rightarrow 2$ optimal cloning machine was designed by
Bu\v{z}ek and Hillery~\cite{Buzek96}. This cloning machine,
referred to as the universal cloning machine (UC), prepares
two approximate copies of an unknown pure qubit state with the
same fidelity $F=5/6$. This means that the UC is state independent
(i.e., the cloning is equally good for any pure qubit state)
and symmetric (i.e., the copies are identical).

The case of the UC producing the infinite number of
copies~\cite{Gisin} allowed to establish the classical limit
of $F=5/6$ for copying quantum information, which corresponds
to the best copying operation achieved by classical operations.

Next, the concept of optimal cloning was extended to include
cloning of qudits, cloning of continuous-variable systems, and
state-dependent cloning (non-universal cloning).

The state-dependent cloning machines can produce clones of a
specific set of qubits with much higher fidelity than
$F=5/6$~\cite{Buzek98,Bruss98a,Niu,Bruss00,F1,F2,
Demkowicz,Fan,Hu,Bartkiewicz,Bartkiewicz2} (see also
reviews~\cite{cloning1,cloning2} and references therein).

The study of state-dependent cloning is well motivated since we
often have some {\it a priori} information about a given quantum
state that we want to clone and by employing the available
information, we can construct a cloning machine which surpasses
the UC for some {\it a priori} specified set of qubits.

For example, if the qubits are taken from the equator of the Bloch
sphere then by using the so-called optimal {\em phase-covariant
cloners} (PCCs)~\cite{Bruss00,Fan}, one achieves higher
fidelity than that for the UC.

The phase-covariant or phase-independent cloning was further
generalized by Fiur\'a\v{s}ek~\cite{F2} who studied the PCCs of
qubits of known expectation value of Pauli's $Z\equiv
\hat{\sigma}_z$ operator and provided two optimal symmetric
cloners: one for the states in the northern and the other for
those in the southern hemisphere of the Bloch sphere.

Further works on phase-independent cloning included cloning of
qubits uniformly distributed on a belt of the Bloch
sphere~\cite{Hu} and {\em mirror phase-covariant cloning}
(MPCC)~\cite{Bartkiewicz}, for qubits of known the modulus of the
expectation value of Pauli's $\hat{\sigma}_z$ operator.

The cloning transformation for the MPCC requires one ancilla and
has the following unitary form in the computational basis:
\begin{eqnarray}\nonumber
|0\>_{\rm in}&\rightarrow& \Lambda|00\>_{1,2}|0\>_{\rm anc}
+\bar{\Lambda}|\psi_+\>_{1,2}|1\>_{\rm anc},
\\
|1\>_{\rm in}&\rightarrow& \Lambda|11\>_{1,2}|1\>_{\rm anc}
+\bar{\Lambda}|\psi_+\>_{1,2}|0\>_{\rm anc},\label{N01}
\end{eqnarray}
where $\Lambda^2 + \bar{\Lambda}^2=1$ and $|\psi_+\>=1/\sqrt{2}
(\ket{01}+\ket{10})$ is one of the Bell states. $\Lambda$
explicitly depends on modulus of expectation value of
$\hat{\sigma}_z$ as follows:
\begin{eqnarray}
\Lambda &=& \sqrt{\frac{1}{2}+
\frac{\cos^2\theta}{2\sqrt{P}}},\label{N15}
\end{eqnarray}
where
\begin{eqnarray}
P\equiv P(\theta)=2-4\cos^2\theta+3\cos^4\theta
\end{eqnarray}
and $\<\hat{\sigma}_z\>=\cos\theta$. The average fidelity $F$ over
the input states of the cloning machine is equal to $F=0.8594$ and
is larger than the fidelity in the case of the UC, which is equal
to $F=0.8333$. Moreover, from Ref.~\cite{Bartkiewicz2} follows
that any optimal cloning machine that copies a phase-covariant set
of qubits and exhibits mirror $xy$-plane symmetry is described by
such general transformation. Therefore, the proposed experimental
setup can be used for cloning an arbitrary set of qubits of the
described symmetry. It is worth noting that former proposals of
realizations of the MPCC in linear-optical
systems~\cite{Bartkiewicz,Bartkiewicz2} and quantum
dots~\cite{Bartkiewicz,Adam2002} were discussed formally without
referring to experimental setups.

In this paper we propose an optical implementation of the
MPCC~\cite{Bartkiewicz} based on a generalized version of the
setup described by \v{C}ernoch \etal~\cite{Cernoch2006} (see also
Ref.~\cite{ExperimentPCC}). The experimental setup can
equally well perform operations of the UC, PCC, and MPCC in
special cases, i.e., for the proper choice of $\Lambda$ (the
explicit formulas can be found in Ref.~\cite{Bartkiewicz2}).

In the following sections we analyze the performance of our setup
accounting for various losses and imperfections such as finite
efficiency of generating a pair of entangled photons in the type-I
spontaneous parametric down conversion (SPDC), the influence of
choosing various parameters of an unbalanced beam splitter
(splitting ratios for vertical and horizontal polarization
components), finite detector efficiency, dark counts, and finite
resolution of applied detectors. For simplicity, we neglect the
effects of mode mismatch on the fidelity of the MPCC. Analysis of
such losses would require application of a pulse-mode formalism
(see, e.g., Ref.~\cite{Sahin2002}).

The paper is organized as follows. In Sec.~II, we present a setup
implementing the optimal symmetric $1\rightarrow 2$ mirror
phase-covariant cloning of a qubit and study the influence of
imperfections of the beam splitter on the performance of cloning.
In Sec.~III, we study the performance of the setup assuming
imperfect photon detectors by means of the positive operator
valued measure (POVM) formalism. We conclude in Sec.~IV.

\section{A proposal for practical photonic implementation of the MPCC}

\subsection{Initialization}

\begin{figure}[t!]
\epsfig{file=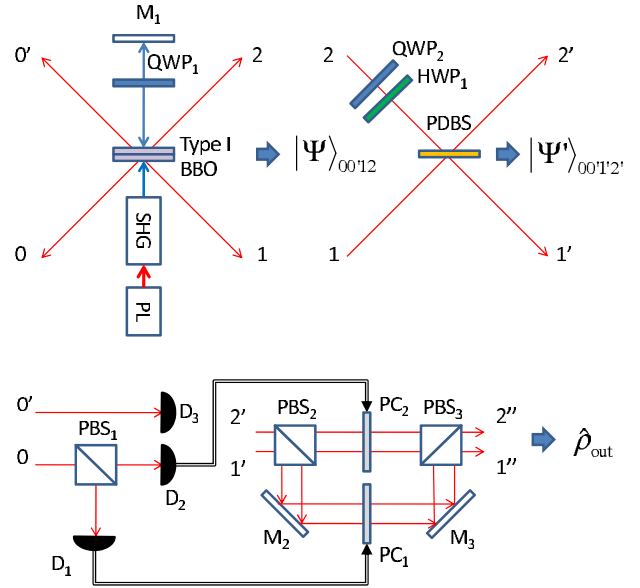,width=9.0cm}\hspace*{0mm} \caption[]{Scheme
of the experimental setup used to implement the MPCC. We use the
following acronyms for the standard optical elements: QWP --
quarter-wave plate, HWP -- half-wave plate, PBS -- balanced
polarizing beam splitter, PDBS -- polarization-dependent beam
splitter for different splitting ratios for H and V polarizations,
D -- detector, PC -- Pockels cell, PL -- pulsed laser, SHG --
second-harmonic generation and M -- mirror. Double solid lines
denote transfer of classical information. The intermediate state
$\ket{\Psi}_{00^\prime 12}$ is prepared by means of spontaneous
parametric down conversion of type-I (see, e.g.,
Ref.~\cite{White1999}) using a stack of two $\beta$-barium borate
(BBO) crystals. Next, $\ket{\Psi}_{00^\prime 12}$ is transformed
into $\ket{\Psi^\prime}_{0,0^\prime 1^\prime 2^\prime}$ by first
setting the input state $\ket{\psi}_2$ to be cloned with the
QWP${}_2$ and HWP${}_1$, and then mixing the modes 1 and 2 on the
PDBS (this can be considered as the first step of the actual
cloning). Finally, $\ket{\Psi^\prime}_{0 0^\prime 1^\prime
2^\prime}$ is subject to classical feedforward, driven by the
measurement outcomes of D${}_1$ and  D${}_2$. As a result, we
obtain state $\hat{\rho}_{out}$ which is the outcome of the
cloning machine as long there is one photon in mode
1${}^{\prime\prime}$ and 2${}^{\prime\prime}$. Detector D${}_3$ is
used as a trigger for the experiment, which practically eliminates
the probability of having vacuum in mode 2. This is due to low
dark-count rate of modern photon detectors.} \label{fig1}
\end{figure}

The initial entangled state is prepared by using parametric down
conversion of the first type (see Refs.~\cite{Kwiat1995,
Kwiat1999, White1999,Rangarajan2009}). The output of a pulsed
laser (PL), with angular frequency $\omega_0$, is frequency
doubled in a nonlinear crystal to produce pulses of ultraviolet
(UV) light of angular frequency $2\omega_0$. The UV pulses are
then used to pump twice (in forward and backward directions) a
pair of nonlinear crystals which are stacked together such that
their optical axes are orthogonal to each
other~\cite{White1999,Tashima}. The crystals are for the type-I
SPDC to produce photon pairs in two modes (idler and signal) of
the same polarization and of half the frequency of the PL. In the
forward pumping direction, the polarization of the UV beam is set
to vertical so that an H-polarized photon pair in modes 2 and
$0^\prime $ are generated. The remaining (non-down-converted)
portion of the UV beam first passes through a quarter wave plate
(QWP$_1$) which changes its polarization into an ellipsoidal
polarization. A mirror M$_1$ placed after the QWP$_1$ reflects
this beam and sends it through the QWP$_1$ again which further
changes the polarization of the beam into diagonal polarization.
This diagonally polarized beam pumps the crystals in the backward
direction creating the entangled photon pair $\ket{\psi_+}=(\ket
{1_{\rm H}}_0 \ket{1_{\rm V}}_{1} +\ket {1_{\rm V}}_0\ket {1_{\rm
H}}_{1})/ \sqrt{2}$. However, the total state of the system in
modes 0, 0${}^\prime$, 1, and 2 is more complex than
\begin{equation}
\ket{\Psi}=\ket{\psi_+}_{0,1}\ket{1_{{\rm H}}}_{0^\prime}
\ket{1_{{\rm H}}}_{2},
\end{equation}
which we use in our further analytical considerations. On the one
hand, the SPDC is probabilistic and the state $\ket{\Psi}$
consists also of the vacuum and higher-order SPDC terms. On the
other hand, for the circuit to work we require fourfold
coincidence count in all modes and a very low dark-count rate of
modern photon detectors (dark count probability of the order of
$10^{-6}$) allows us to effectively eliminate the vacuum state
from the further considerations. Moreover, the measurement of a
photon in mode 0 is polarization dependent, which is further used
in the feedforward processing. So, finally the system is prepared
in the state
\begin{eqnarray}\nonumber
\ket{\Psi}&=&\mathcal{N} \left[\gamma^2
\E^{2i\phi}\ket{\psi_+}_{01}\ket{1_{{\rm
H}}}_{0^\prime}\ket{1_{{\rm H}}}_{2}\right.\\ \nonumber
&&+ \gamma^3 \E^{3i\phi}\left(\ket{\psi_+}_{01}\ket{2_{{\rm H}}}_{0^\prime}\ket{2_{{\rm H}}}_{2}\right.\\
&&+ \left.\left. \ket{\epsilon}_{01}\ket{1_{{\rm
H}}}_{0^\prime}\ket{1_{{\rm H}}}_{2}\right) +
\mathcal{O}(\gamma^4)\right],\label{psi}
\end{eqnarray}
where $$\ket{\epsilon}_{01}= \frac12(
 \ket{1_{{\rm H}}1_{{\rm V}}}_{0} +
 \ket{1_{{\rm H}}1_{{\rm V}}}_{1} +
 \ket{2_{{\rm H}}}_{0}\ket{2_{{\rm V}}}_{1} +
 \ket{2_{{\rm V}}}_{0}\ket{2_{{\rm H}}}_{1}).$$
Moreover, $\gamma$  describes the efficiency of the SPDC and
depends on the amplitude of the incident field and properties of
the nonlinear crystal, $\phi$ is the phase shift caused by the
SPDC, and $\mathcal{N}$ is the normalization constant. Typically
$\gamma^2 = 0.01$~\cite{Tashima}, so for simplicity we can neglect
the terms of amplitudes of order higher than $\gamma^3$ since
probability of the occurrence of such event is very low as
$|\mathcal{O}(\gamma^4)|^2 = \mathcal{O}(\gamma^8)$. We will
consider a more complete form of $\ket{\Psi}$ only in Sect.~III.

Next, we prepare the arbitrary state to be cloned in mode 2 by a
combination of a half-wave plate (HWP$_1$) and QWP$_2$. The input
state is passed into mode 2. It is given in the following form:
\begin{equation}\label{input}
|\psi\>_{{\rm 2}}=(\alpha\abh+\beta\abv)|0\>_{{\rm 2}},
\end{equation}
where $\alpha = \cos(\theta/2)$ and $\beta =
\E^{i\delta}\sin(\theta/2)$. Later, modes 1 and 2 are mixed on an
unbalanced polarization-dependent beam splitter (PDBS). The PDBS
transforms the input in the following way
\begin{eqnarray} \nonumber
\aah &\rightarrow& \sqrt{1-\mu}\ach - \sqrt{\mu}\adh\;,
\\ \nonumber
\aav &\rightarrow& \sqrt{1-\nu}\acv + \sqrt{\nu}\adv\;,
\\ \nonumber
\abh &\rightarrow& \sqrt{\mu}\ach + \sqrt{1-\mu}\adh\;,
\\
\abv &\rightarrow& \sqrt{\nu}\acv - \sqrt{1-\nu}\adv\;.
\end{eqnarray}
The MPCC can be implemented when
\begin{equation}
\mu + \nu  =1.
\end{equation}
The most convenient situation is when $\mu=\mu_0=(1-1/\sqrt{3})/2$
and $\nu=\nu_0=(1+1/\sqrt{3})/2$, i.e., $1-2\mu=2\nu-1
=\sqrt{2\mu\nu}=1/\sqrt{3}$. Analogical conditions for the PCC
were given by Fiur\'a\v{s}ek~\cite{F2}. Finally, the state of the
system after the action of the PDBS (for $\mu=\nu$) is given by
the following expression:
\begin{eqnarray}
\nonumber |\Psi'\>&=& {\cal N}\,'\left[
{\alpha}\azv\left(\sqrt{\mu\nu}(\ach\ach-\adh\adh)\right.\right.
\\ \nonumber
&&\left.+(1-2\mu)\ach\adh\right)
\\ \nonumber
&&+{\beta}\azv\left(\nu\ach\acv + \mu\adh\adv\right.
\\ \nonumber
&&\left.-\sqrt{\mu\nu}(\ach\adv+\acv\adh) \right)
\\ \nonumber
&&+{\alpha}\azh\left(\mu\ach\acv + \nu\adh\adv \right.
\\ \nonumber
&&\left.+\sqrt{\mu\nu}(\acv\adh+\ach\adv)\right)
\\ \nonumber
&&+{\beta}\azh\left(\sqrt{\mu\nu}(\acv\acv -\adv\adv)\right.
\\
&&\left.\left.+(1-2\mu)\acv\adv\right)\right] \abph|0\>_{{\rm
00'1'2'}},
\end{eqnarray}
where ${\cal N}\,'$ is a normalization constant.

\subsection{Feedforward}
In order to implement the MPCC we also apply a feedforward
technique (see Refs.~\cite{Bohi,Prevedel2007}), i.e., photons of
the same polarization as detected in mode 0 are damped in modes 1'
and 2'. The element implementing the damping is based on a Pockels
cell and two PBSs and is presented in Fig.~1. As it was shown in
Ref.~\cite{Bohi} that such operation can be performed with high
fidelity of more than 99\%. The final density matrix of the system
is given as
\begin{equation}
\hat{\rho}_{\rm out}\!=\! \tr_{\rm
00'}\big[\big(\hat{\Pi}_{1H}^{0}\hat{\Pi}_{0V}^{0}\hat{D}_{H} \hat
\rho'\hat{D}_{H}^\dagger + \hat{\Pi}_{1V}^{0}\hat{\Pi}_{0H}^{0}
\hat{D}_{V}\hat \rho' \hat{D}_{V}^\dagger
\big)\hat{\Pi}_{1}^{0'}\big],\label{rho_out}
\end{equation}
where $\hat \rho' = |\psi'\>\<\psi'|$ is the output the state
after the action of the unbalanced PDBS,
$\hat{D}_{H}=\hat{\Gamma}_{{\rm1'H}}\hat{\Gamma}_{{\rm 2'H}}$,
$\hat{D}_{V}=\hat{\Gamma}_{{\rm1'V}}\hat{\Gamma}_{{\rm 2'V}}$,
where  $\hat \Gamma_{iV}$ ($\hat \Gamma_{iH}$) is the operation
acting on photons in the $i$th spatial mode, which corresponds to
the conditional application of a Pockels cell (see Fig.~1).
$\hat{\Pi}_{jH}^{i}$ ($\hat{\Pi}_{jV}^{i}$) are the POVM operators
describing the probability of detection of the $j$ H(V)-polarized
photons in the $i$th mode. The damping operation can is described
as
\begin{eqnarray}
\nonumber \hat{\Gamma}_{i,s}\ket{m_{\rm H},n_{\rm V}}_j&=&
\left[\delta_{ij}(-\gamma)^{n\delta_{s,{\rm V}}}
\gamma^{m\delta_{s,{\rm H}}}\right.\\
&&\left.+(1-\delta_{ij})\right]\ket{m_{\rm H},n_{\rm V}}_j,
\end{eqnarray}
where $\delta_{ij}$ is Kronecker's delta, $i$ and $j$
enumerate the spatial modes 1${}^\prime$ and 2${}^\prime$, while
$r$ and $s$ stand for polarization H and V, respectively.
Moreover, $\gamma$ is a damping parameter, which in the case of a
perfect PDBS is equal to $\gamma=\bar{\Lambda}/\Lambda$.

\subsection{Post-selection}

\begin{figure}[t!]
\epsfig{file=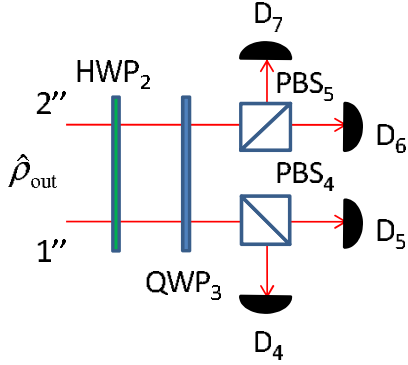,width=6.0cm}\hspace*{0mm} \caption[]{The
final part of the setup that can be used for verification of the
cloning operation. The HWP${}_2$ and QWP${}_3$ rotate $\ket{\psi}$
and $\ket{\overline{\psi}}$ to $\ket{1_H}$ and $\ket{1_V}$.
Cloning is successful when two of the detectors click, one of the
pair (D${}_4$,D${}_5$) and one of the pair (D${}_6$,D${}_7$). Four
detectors are used in order to evaluate the fidelity of the
cloning operations as given in Eq.~(\ref{FidC}), where the pairs
of detectors (D${}_4$,D${}_5$) and (D${}_6$,D${}_7$) correspond to
the POVMs ($\hat{\overline{\Pi}}_1$, $\hat{\Pi}_1$) and
($\hat{\Pi}_1$,$\hat{\overline{\Pi}}_1$), respectively.}
\label{fig2}
\end{figure}
The cloning procedure is successful as long as there is only one
photon in every outgoing mode. Probability of the coincidence
count (probability of success) is given by the following
expression:
\begin{equation}
P_{\rm success}=\tr_{3,4}(\hat{\rho}_{{\rm
out}}\hat{\Pi}_{1}^{1^{\prime\prime}}\hat{\Pi}_{1}^{2^{\prime\prime}})
= \frac{1}{6\Lambda^2},
\end{equation}
where  $1/\sqrt{2}\leq\Lambda\leq 1$ for the MPCC. Hence, the
probability of successful cloning $P_{\rm success}$ varies from
$1/3$ to $1/6$ (given that we work with perfect detectors and a
perfect source of entangled photons) depending on the states we
want to clone in an optimal way. The proposed implementation is
probabilistic, but the probability of successful cloning is much
higher than in the case of using the simple quantum circuit
proposed in Ref.~\cite{Bartkiewicz}, which requires four
controlled NOT (CNOT) gates, with the best known nondestructive
optical CNOT gates with the success rate of $1/4$ of
Pittman~\cite{Pittman2001} (for a review see
Ref.~\cite{Bartkowiak2010}). The optimal cloner constructed in
such way will have the success rate of $1/256$.

\subsection{Fidelity of the proposed experimental setup}

\begin{figure}[]
\hspace*{64mm}(a)
 \epsfxsize=8.5cm\epsfbox{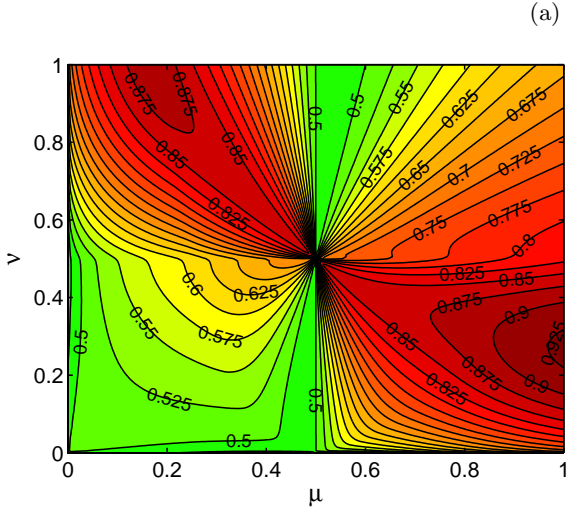}\hspace*{0mm}\\
\hspace*{64mm}(b)
\epsfxsize=8.5cm\epsfbox{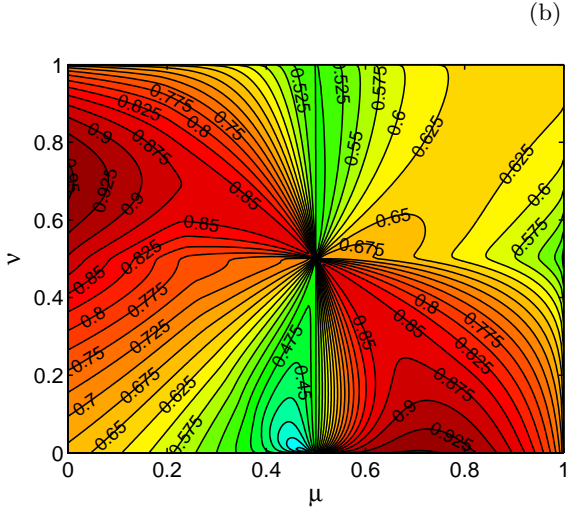}\vspace{0mm}\\
\hspace*{64mm}(c)
 \epsfxsize=8.5cm\epsfbox{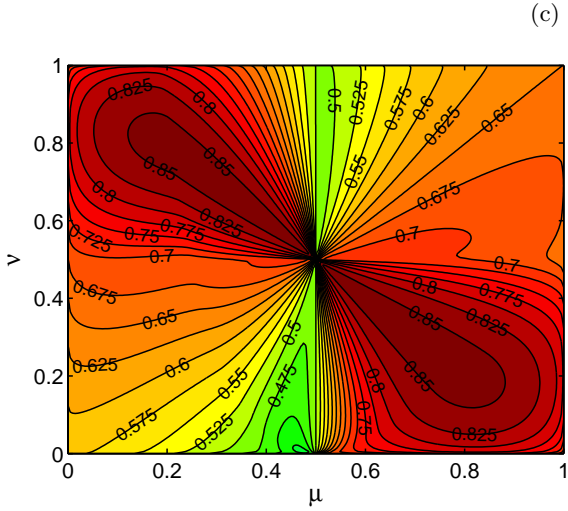}\hspace*{0mm}
\caption[]{Comparison of the average cloning fidelities of the
first clone (a), the second clone (b) and its average over the two
clones (c) for the MPCC. The average fidelity over the two clones
reaches its maximum $F=0.8594$ for $\nu = 1-\mu$ and $\mu$ given
by Eq.~(\ref{mu}). We observe that the cloning fidelity is
symmetric for the parameters close to $\nu = 1-\mu$. The areas of
high fidelity ($>0.85$) are large in all three cases. Thus, the
setup is robust to variations of $\mu$ or $\nu$.} \label{fig3}
\end{figure}

\begin{figure}[]
 \epsfxsize=8.5cm\epsfbox{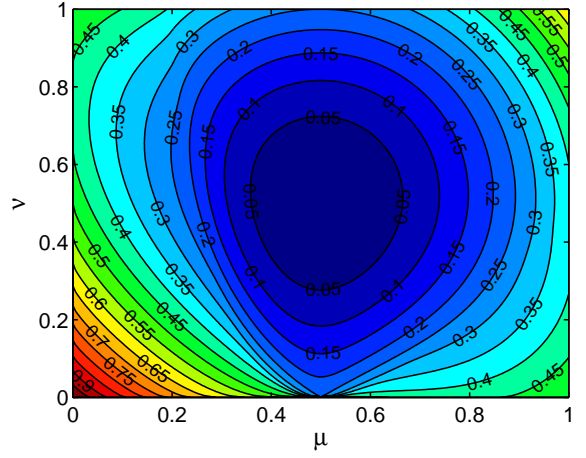}\vspace{0mm}
\caption[]{The average success probability of the proposed setup
in the case of the MPCC, which corresponds to the probability of
finding one photon both in modes 1${}^{\prime\prime}$ and
2${}^{\prime\prime}$. The probability of successful coincidence
count in both modes increases radially from the center of the
figure where it reaches 0. We find the best $\mu$ and $\nu$ by
finding such conditions for which the average fidelity and the
probability of success are simultaneously maximized (see Fig.~3).
This happens when $\nu = 1-\mu$ and the inequality given in
Eq.~(\ref{mu}) is saturated. } \label{fig4}
\end{figure}

In order to describe the quality of the cloning we use single-copy
fidelity
\begin{equation}
F_i=\frac{\<\psi|\tr_{3-i} \hat{\rho}_{{\rm
out}}|\psi\>}{\tr\hat{\rho}_{{\rm out}}}.
\end{equation}

However, to describe the overall performance of a cloning machine
it is more convenient to use the average single-copy fidelity
\begin{equation}
F=\frac12\int_0^{2\pi} {\rm d}\phi \int_0^\pi {\rm d}\vartheta\, g
(\vartheta,\phi) [F_1(\vartheta,\phi)+F_2(\vartheta,\phi)]
\label{N04}
\end{equation}
which is an average over all possible input qubits defined by the
distribution function $g (\vartheta,\phi)$. In case of the MPCC,
$g$ distribution form Eq.~(\ref{N04}) is given by
\begin{equation}
g_{\theta}(\vartheta,\phi) =\frac1{4\pi}[\delta(\vartheta-\theta)
+\delta(\vartheta+\theta-\pi)], \label{N07}
\end{equation}
in terms of Dirac's $\delta$-function. Moreover, we added
subscript $\theta$ to indicate {\em a priori} knowledge about the
input state.

One can easily check that the resulting expression for the average
single-copy fidelity is the same as for the
MPCC~\cite{Bartkiewicz} and given by
\begin{equation} \label{FidMPCC}
F\equiv F_1=F_2=\frac{1+\Lambda^2}{2}-\frac{1}{2}
\Lambda\left(\Lambda-\bar{\Lambda}\sqrt{2}\right)\sin^2\theta.
\end{equation}
From Eq.~(\ref{FidMPCC}) follows that the average cloning fidelity
over all possible input states of the MPCC (over all $\theta$ -
the average is over all possible circles and their
mirror-symmetric counterparts) is $F = 0.8594$.

Note that for simplicity of exposition, we focus here on the
MPCC, which is the simplest nontrivial example of cloning of
mirror-symmetric distributions $g(\theta)$ on Bloch sphere, where
$g(\theta)$ is a sum of two Dirac's $\delta$-functions. However,
in the case of other mirror-symmetric phase-covariant qubit
distributions we obtain different values the average
cloning fidelity and success rate of the proposed experimental
setup. For example, in the case of the UC we obtain $F = 0.8333$
(the average is over the whole Bloch sphere), and $F = 0.8536$
(the average over the equator of Bloch sphere) in the case of the
PCC. For the UC and PCC, we have $\Lambda=\sqrt{2/3}$ and
$\Lambda=1/\sqrt{2}$, respectively.

\section{Practical considerations for experimental implementation}

\subsection{Choosing the parameters of unbalanced
polarization-dependent beam splitter}
In order to perform the required quantum transformation in some
cases one needs to use a polarization-dependent beam splitter of
some strictly chosen values of reflectance or transmittance (see
Refs.~\cite{Tashima,Cernoch2006}). However, there are no perfect
polarization beam  splitters (see
Refs.~\cite{Cernoch2006,Tashima}). In practice one can apply some
mechanisms to compensate for the imperfections of the beam
splitter (see Refs.~\cite{Cernoch2006}). In our case we use
feedforward and obtain less strict requirements on optical
realization of the symmetric covariant cloner than stated in
Ref.~\cite{Cernoch2006}, where ${1-\nu}=1/2(1+1/\sqrt{3})$ and
${1-\mu}=1/2(1-1/\sqrt{3})$ must have fixed values.

In the proposed experimental realization, it is enough that
$\mu+\nu=1$ is satisfied. Otherwise, the single copy fidelity
drops and cloning is no longer symmetric (see Fig.~3). Given that
$\mu+\nu=1$ is satisfied, it is enough that the damping parameter
\begin{equation}
\gamma=\frac{\bar{\Lambda}(1-2\mu)}{\Lambda\sqrt{2\mu\nu}}
\end{equation}
and the fidelity of a single clone are the same as in the perfect
case. (Please note that only $|\lambda|\leq1$ is physical). Hence,
imperfections of the PDBS (given that $\mu+\nu=1$) result only in
decreasing the success rate of the setup (see Fig.~4). Therefore,
the probability of successful cloning is given by the following
expression:
\begin{equation}
P_{\rm success} = \frac{(1-2\mu)^2}{2\Lambda^2},
\end{equation}
where
\begin{equation}\label{mu}
\frac12 \left(1-\frac{1}{\sqrt{3}}\right)\leq\mu\leq
\frac12\left(1 +\frac{1}{\sqrt{3}}\right).
\end{equation}

\subsection{Influence of detector imperfections}

Detectors play an important part in the proposed experiment. As
one can see in Eq.~(\ref{rho_out}), the density matrix
$\hat{\rho}_{\rm out}$ depends explicitly on the measurements
performed on the ancillary qubits. Also in practical realizations
of the cloning machine, the fidelity of the cloning process can be
evaluated by measuring polarization of photons in modes 1$''$ and
2$''$ in the basis of $\ket{\psi}$ and $\ket{\overline{\psi}}$ as
described in Ref.~\cite{Cernoch2006}. This gives seven detectors
in total, however we analyze only the cases when four detectors
(one photon per detector) click at the same time. For simplicity,
we assume that all the detectors are characterized by the same
parameters.

There are two basic types of photon detectors that can be used in
the experiment: single-photon counters and ON/OFF detectors. Since
we cannot exclude completely the possibility of higher-order SPDC
events [see Eq.~(\ref{psi})] we investigate the implications of
using both types of the detectors.

\subsubsection{Single-photon counters}

First we analyze single-photon counters, which can discriminate
between vacuum, detection one photon, and detection of many
photons. We describe imperfections of these detectors by the
following POVM operators~\cite{Sahin2001}:
\begin{eqnarray}
\hat{\Pi}_{0}&=& \sum_{m=0}^\infty e^{-\zeta}(1-\eta)^{m}|m\rangle
\langle m|,
\nonumber\\
\hat{\Pi}_{1}&=& \sum_{n=0}^1 \sum_{m=n}^\infty
e^{-\zeta}\zeta^{1-n}\eta^{n}m^{n}(1-\eta)^{m-n}|m\rangle \langle
m|,
\nonumber\\
\hat{\Pi}_{N\geq2}&=& \hat{\openone}-\hat{\Pi}_{0}-\hat{\Pi}_{1}\,
. \label{N19}
\end{eqnarray}
where $\eta$ is quantum efficiency of the detectors and $\zeta$
stands for the dark-count rate (typically of the order of
$~10^{-6}$).

\subsubsection{ON/OFF detectors}

We also analyze the ON/OFF detectors (also referred to as
conventional or bucket detectors), which can discriminate only
between vacuum and any other number of photons. The difference
between the single-photon counters and ON/OFF detectors is
negligible in the case of low dark-count rate. Since we are
interested only in such events where the number detector
``clicks'' is equal to the assumed number of photons in the
system. We use the following POVM operators~\cite{Sahin2001}:
\begin{eqnarray}
\hat{\Pi}_{0} &=& \sum_{m=0}^\infty
e^{-\zeta}(1-\eta)^{m}|m\rangle \langle m|,
\nonumber\\
\hat{\Pi}_{N\geq1} &=&  \hat{\openone}-\hat{\Pi}_{0}\,
.\label{N18}
\end{eqnarray}

\subsubsection{Expected fidelity and probability of cloning}

In our proposed experimental setup we use post-selection, thus the
average fidelities of the clones can be expressed via coincidences
as~\cite{Cernoch2006}:
\begin{eqnarray}
 F_1= \frac{C_{11} + C_{10}}{P_{\rm success}},
\quad
 F_2= \frac{C_{11} + C_{01}}{P_{\rm success}},
\label{FidC}
\end{eqnarray}
where
\begin{eqnarray}
P_{\rm success} = C_{00} + C_{01} + C_{10} + C_{11}
\end{eqnarray}
is the probability of success (successful post-selection) and
$C_{ij}$ ($i,j\in\{0,1\}$) are the following coincidences
\begin{eqnarray}\nonumber
C_{11}=\tr\left(\hat{\rho}_{{\rm out}}{\hat{\Pi}}_1 \otimes
{\hat{\Pi}}_1\right),~ C_{10}=\tr\left(\hat{\rho}_{{\rm
out}}{\hat{\Pi}}_1 \otimes \hat{\overline{\Pi}}_1\right),
\\\nonumber
C_{01}=\tr\left(\hat{\rho}_{{\rm out}}\hat{\overline{\Pi}}_1
\otimes {\hat{\Pi}}_1\right),~ C_{00}=\tr\left(\hat{\rho}_{{\rm
out}}\hat{\overline{\Pi}}_1 \otimes \hat{\overline{\Pi}}_1\right).
\end{eqnarray}
Here, the POVMs $\hat{\Pi}_1$ and $\hat{\overline{\Pi}}_1$
correspond to the detection of a photon in the state $\ket{\psi}$
and $\ket{\overline{\psi}}$, respectively. As one can see in
Eq.~(\ref{rho_out}), $\hat{\rho}_{{\rm out}}$ depends on the
quality and type of the photon detectors. Moreover, it also
depends on the efficiency of generation of the entangled photon
pairs [see Eq.~(\ref{psi})]. In the case of perfect detectors
(both single-photon counters and ON/OFF detectors) we have
$\hat{\Pi}_1 = \ket{1_\psi}\bra{1_\psi}$. The influence of
imperfections of measurements on the fidelity of cloning and
success rate for single-photon counters (ON/OFF detectors) is
summarized in Tables~I and~II.

\begin{table}[!ht]
\caption{The influence of imperfections of the detectors
(neglecting dark counts) on the average success rate $P_{\rm
success}$ and the average fidelities ${F}_1$ and ${F}_2$ of two
clones, where $\eta$ is detector's efficiency (some achievable
values can be found in Refs.~\cite{Yamamoto2001,
Takeuchi1999}). The results show that the proposed cloning machine
is essentially robust to finite-efficiency and finite-resolution
of detectors. The loss of the fidelity caused by the imperfections
is less than 1\%. However, the theoretical limit of the maximal
cloning fidelity can be reached only in the case of photon-number
discriminating detectors.} \vspace{.3cm}
\begin{ruledtabular}
\begin{tabular}{c|ccc|ccc}
$\eta$  & $P_{\rm success}$\footnotemark[1]  & ${F}_1$\footnotemark[1] & ${F}_2$\footnotemark[1] & $P_{\rm success}$ \footnotemark[2] & ${F}_1$\footnotemark[2] & ${F}_2$\footnotemark[2]\\
\hline
1.00 &   0.2552  & 0.8594   & 0.8594  &   0.2598  & 0.8567  & 0.8569     \\
0.90 &   0.1357  & 0.8591   & 0.8592  &   0.1387  & 0.8562  & 0.8564     \\
0.80 &   0.0671  & 0.8588   & 0.8589  &   0.0688  & 0.8555  & 0.8558     \\
0.70 &   0.0302  & 0.8583   & 0.8584  &   0.0311  & 0.8548  & 0.8551     \\
0.60 &   0.0120  & 0.8576   & 0.8578  &   0.0124  & 0.8540  & 0.8543     \\
0.50 &   0.0041  & 0.8567   & 0.8569  &   0.0042  & 0.8531  & 0.8534     \\
0.40 &   0.0011  & 0.8555   & 0.8558  &   0.0011  & 0.8521  & 0.8524     \\
0.30 &   0.0002  & 0.8540   & 0.8543  &   0.0002  & 0.8510  & 0.8513     \\
\end{tabular}
\end{ruledtabular}
\footnotetext[1]{Single-photon counters.} \footnotetext[2]{ON/OFF
detectors.} \label{summary1}
\end{table}

\begin{table}[!ht]
\caption{The influence of finite dark-count rate of the detectors
assuming perfect efficiency on the average success rate $P_{\rm
success}$ and the average fidelities ${F}_1$ and ${F}_2$ of two
clones, where $\zeta$ is the detector's dark-count rate. The
influence of the dark counts for the usual dark-count rates
($\zeta$ of the order of $10^{-6}$~\cite{Sahin2001}) is
negligible. For both types of the detectors the setup is robust
(less than 1\% drop of the average fidelity) up to $\zeta$ of the
order of 0.001. It is seen that probability of coincidence count
increases with $\zeta$ for the ON/OFF detectors and drops in the
case of the photon-number discriminating detectors (single-photon
counters). The ON/OFF detectors register false successful events
as true coincidences. The single-photon counters are better in the
case of low dark-count rates (most of practical situations), but
for $\zeta>0.0001$ we observe that the performance of the machine
is better when the ON/OFF detectors are applied.} \vspace{.3cm}
\begin{ruledtabular}
\begin{tabular}{c|ccc|ccc}
$\zeta$  & $P_{\rm success}$\footnotemark[1]  & ${F}_1$\footnotemark[1] & ${F}_2$\footnotemark[1] & $P_{\rm success}$ \footnotemark[2] & ${F}_1$\footnotemark[2] & ${F}_2$\footnotemark[2]\\
\hline
$10^{-6}$   & 0.2552 & 0.8594 & 0.8594   &  0.2598   & 0.8567   &  0.8569   \\
$10^{-5}$   & 0.2552 & 0.8594 & 0.8594   &  0.2598   & 0.8567   &  0.8569   \\
$10^{-4}$   & 0.2550 & 0.8589 & 0.8589   &  0.2598   & 0.8566   &  0.8568   \\
$10^{-3}$   & 0.2536 & 0.8543 & 0.8543   &  0.2600   & 0.8557   &  0.8559   \\
$10^{-2}$   & 0.2403 & 0.8094 & 0.8094   &  0.2620   & 0.8470   &  0.8472   \\
$10^{-1}$   & 0.1409 & 0.4724 & 0.4724   &  0.2718   & 0.7818   &  0.7820   \\
\end{tabular}
\end{ruledtabular}
\footnotetext[1]{Single-photon counters.} \footnotetext[2]{ON/OFF
detectors.} \label{summary2}
\end{table}

\section{Applicability to arbitrary mirror-symmetric phase-covariant cloning}

For simplicity, so far we analyzed the setup for the
MPCC~\cite{Bartkiewicz} only. However, the general cloning
transformation given in Eq.~(\ref{N01}) is optimal for cloning of
arbitrary mirror-symmetric distributions on Bloch sphere.

Recently we showed~\cite{Bartkiewicz2} that the optimal symmetric
$1\rightarrow 2$ of an arbitrary axisymmetric distribution of
qubits $g(\theta)$ (distribution of expectation values
$\<\hat{\sigma}_z\>=\cos\theta$ for a set of qubits). Any
$g(\theta)$ can be expanded in the basis of Legendre polynomials
$P_n(\cos{\theta})$~\cite{Kaplan} as
\begin{eqnarray}
g(\theta) &=& \frac{1}{4\pi} \sum_{n=0}^{\infty}(2n+1)a_n P_n(\cos{\theta}),\\
a_n  &=& \int_{0}^{2\pi}\int_{-1}^{1}
g(\theta)P_n(\cos{\theta})\,{\rm d}\cos{\theta}\,{\rm d}\phi.
\label{an}
\end{eqnarray}
In Ref.~\cite{Bartkiewicz2} we showed that the optimal cloning
transformation depends only on first three terms of this
expansion. Moreover, for a normalized ($a_0=1$) mirror symmetric
(invariant to the action of discrete Weyl-Heisenberg group)
distribution we obtain $a_1=0$. Such case includes as special
cases the PCC for $\theta=\pi/2$, the MPCC~\cite{Bartkiewicz}, and
the UC of Bu\v{z}ek and Hillery~\cite{Buzek96}.

By comparing the results from Ref.~\cite{Bartkiewicz} with those
from~\cite{Bartkiewicz2} we find that $\bar{\Lambda}$ from
Eq.~(\ref{N01}) in general depends on a single parameter as
follows:
\begin{equation}
\Lambda =
\sqrt{\frac{1}{2}+\frac{1}{2}\sqrt{1-\frac{8(1-a_2)^2}{3(3+4a_2^2-4a_2)}}}.
\label{Lambda_general}
\end{equation}
Thus, by using appropriate functional form of $\Lambda$ we can
implement various optimal cloning machines such as the PCC, MPCC
and UC with the same experimental setup. Note that for the UC
$a_2=0$ and for the PCC $a_2=-1/2$, i.e., $\Lambda=\sqrt{2/3}$ and
$\Lambda=1/\sqrt{2}$, respectively.

\section{Conclusions}

W investigated experimentally-feasible optimal mirror
phase-covariant cloning, i.e., optimal cloning of arbitrary sets
of qubits of known modulus of expectation value of Pauli's
$\hat{\sigma}_z$ operator. Our definition of the mirror
phase-covariant cloning includes in special cases the universal
cloning (corresponding to cloning of a uniform distribution of
qubits on Bloch sphere) and the phase-covariant cloning (cloning
of equatorial qubits). By identifying the class of
mirror-symmetric phase-covariant distributions of qubits as
subclass of axisymmetric distributions, for which optimal cloning
transformations were obtained in Ref.~\cite{Bartkiewicz2}, we
showed that the cloning transformation we implemented is optimal.

We proposed an optical realization of optimal quantum mirror
phase-covariant $1\to2$ cloning of a qubit, for which the mean
probability of successful cloning varies from $1/6$ to $1/3$
depending on the prior information on the set of qubits to be
cloned. The qubits are represented by polarization states of
photons generated by spontaneous parametric down-conversion of the
first type. The scheme is based on the interference of two photons
on a beam splitter with different splitting ratios for vertical
and horizontal polarization components and additional application
of feedforward by means of Pockels cells.

The phase-covariant cloning machine implemented by \v{C}ernoch
\etal~\cite{Cernoch2006} is less general as it does not include
feedforward that allows to use the setup in cases other than
implementation of the PCC. Moreover, we showed that the
feedforward also allows using a wider range of splitting ratios of
the polarization-dependent beam splitter than in the schemes
without feedforward.

The experimental feasibility of the proposed setup was studied
including various kinds of losses: (i) finite efficiency of
generating a pair of entangled photons in the type-I spontaneous
parametric down conversion, (ii) the influence of choosing various
splitting ratios of an unbalanced beam splitter, (iii) the
use of conventional (ON/OFF detectors) and single-photon
discriminating detectors, (iv) finite detector efficiency, and
(iv) dark counts.

For simplicity, we studied the experimental feasibility of our
setup implementing only the standard MPCC, i.e., which corresponds
to cloning distribution $g(\theta)$ described by two Dirac's
$\delta$-functions. Such analysis can be easily extended to show
the feasibility of our setup for the optimal cloning of arbitrary
distributions $g(\theta)$ that are mirror-symmetric on Bloch
sphere.

We showed that the cloning machine is robust, its fidelity is
expected to be very close to the theoretical limit and is expected
to stay unaffected by the imperfections of the particular elements
other than the PDBS. Robustness of the proposed experimental setup
was confirmed by investigation of influence of the mentioned
imperfections on the average fidelity of clones and success
probability of the MPCC.

Both the success rate and average cloning fidelity were estimated
by means of simplified qubit tomography setup~\cite{James2001}. In
our case, similarly as \v{C}ernoch \etal~\cite{Cernoch2006}, we do
not need to use the complete tomography to determine the fidelity
of the clones (since we {\em a priori} know the input state to
some extent).

The probability of successful cloning is high if compared the
logical circuit described in Ref.~\cite{Bartkiewicz} with all the
CNOT operations replaced with the best optical gates. The setup
proposed in this paper is not only suitable for the MPCC, but also
for any optimal cloning of an arbitrary set of qubits of the axial
and mirror $xy$ symmetry~\cite{Bartkiewicz2} including the
universal, phase-covariant and mirror-phase covariant cloning.

\begin{acknowledgments}
We acknowledge support from the Polish Ministry of Science and
Higher Education under Grants No.~2619/B/H03/2010/38
and~3271/B/H03/2011/40.
\end{acknowledgments}



\begin{thebibliography}{\mode}

\bibitem{Zurek}
W.K. Wootters and W.H. Zurek, Nature (London) \textbf{299}, 802
(1982).

\bibitem{Dieks} D. Dieks, \pla \textbf{92}, 271 (1982).

\bibitem{Buzek96}
V. Bu\v{z}ek and M. Hillery, \pra \textbf{54}, 1844 (1996).

\bibitem{Duan98}
L. M. Duan and G. C. Guo, \prl \textbf{80}, 4999 (1998).

\bibitem{Fuchs97}
C. A. Fuchs, N. Gisin, R. B. Griffiths, C.-S. Niu, and A. Peres,
\pra {\bf 56}, 1163 (1997); D. Bru\ss{}, \prl {\bf 81}, 3018
(1998); H. Bechmann-Pasquinucci and N. Gisin, \pra \textbf{59},
4238 (1999); L. Gyongyosi and S. Imre, WSEAS Trans. Commun. {\bf
9}, 165 (2010).

\bibitem{Bruss01}
D. Bru\ss{}, J. Calsamiglia, and N. L\"utkenhaus, \pra
\textbf{63}, 042308 (2001).


\bibitem{Galvao00}
E. F. Galvao and L. Hardy, \pra \textbf{62}, 022301 (2000).

\bibitem{Experiment}
A. Lamas-Linares, C. Simon, J. C. Howell, D. Bouwmeester, Science
\textbf{296}, 5568 (2002); Z. Zhao \etal, \prl \textbf{95}, 030502
(2005); F. Sciarrino, V. Secondi, and F. De Martini, \pra
\textbf{73}, 040303(R) (2006); M. Sabuncu, U. L. Andersen, and G.
Leuchs, \prl \textbf{98}, 170503 (2007); E. Nagali, D. Giovannini,
L. Marrucci, S. Slussarenko, E. Santamato, and F. Sciarrino, \prl
\textbf{105}, 073602 (2010).

\bibitem{ExperimentPCC2}
J. Du \etal, \prl \textbf{94}, 040505 (2005); H. Chen, X. Zhou, D.
Suter, and J. Du, \pra \textbf{75}, 012317 (2007).


\bibitem{Cernoch2006}
A. \v{C}ernoch, L. Bart{\accent23 u}\v{s}kov\'{a}, J. Soubusta, M.
Je\v{z}ek, J. Fiur\'{a}\v{s}ek, M. Du\v{s}ek, \pra \textbf{74},
042327 (2006).


\bibitem{ExperimentPCC}
L. Bart{\accent23 u}\v{s}kov\'{a}, M. Du\v{s}ek, A. \v{C}ernoch,
J. Soubusta, and J. Fiur\'a\v{s}ek, \prl \textbf{99}, 120505
(2007); J. Soubusta, L. Bart{\accent23 u}\v{s}kov\'{a}, A.
\v{C}ernoch, J. Fiur\'a\v{s}ek, M. Du\v{s}ek, \pra \textbf{76},
042318 (2007); J. Soubusta, L. Bart{\accent23 u}\v{s}kov\'{a}, A.
\v{C}ernoch, M. Du\v{s}ek, and J. Fiur\'a\v{s}ek, \pra
\textbf{78}, 052323 (2008).


\bibitem{ExperimentQD}
A. Zhu, K. H. Yeon, and S. C. Yu, \jpb \textbf{42}, 235501 (2009);
B.-Q. Sun, X.-Q. Shao, A-D. Zhu, K.-H. Yeon, and S.-C. Yu, Phys.
Scr. \textbf{82}, 045006 (2010).


\bibitem{Gisin}
N. Gisin and S. Massar, \prl \textbf{79}, 2153 (1997).

\bibitem{Buzek98}
V. Bu\v{z}ek and M. Hillery, \prl \textbf{81}, 5003 (1998).

\bibitem{Bruss98a}
D. Bru\ss{}, D.P. DiVincenzo, A. Ekert, C.A. Fuchs, C.
Macchiavello, and J.A. Smolin, \pra \textbf{57}, 2368 (1998).

\bibitem{Niu}
C.-S. Niu and R. B. Griffiths, \pra \textbf{58}, 4377 (1998).

\bibitem{Bruss00}
D. Bru\ss{},  M. Cinchetti, G.M. D'Ariano, and C. Macchiavello,
\pra \textbf{62}, 012302 (2000).

\bibitem{F1}
J. Fiur\'a\v{s}ek, \pra \textbf{64}, 062310 (2001).

\bibitem{F2}
J. Fiur\'a\v{s}ek, \pra \textbf{67}, 052314 (2003).

\bibitem{Demkowicz}
R. Demkowicz-Dobrza\'nski, M. Ku\'s, and K. W\'odkiewicz, \pra
\textbf{69}, 012301 (2004).


\bibitem{Fan}
H. Fan, H. Imai, K. Matsumoto, and X. B. Wang, \pra \textbf{67},
022317 (2003).


\bibitem{Hu}
J.Z. Hu, Z.W. Yu, and X.B. Wang, Euro. Phys. J. D \textbf{51}, 381
(2009).

\bibitem{Bartkiewicz}
K. Bartkiewicz, A. Miranowicz, and \c{S}.K. \"Ozdemir, \pra
\textbf{80}, 032306 (2009).

\bibitem{Bartkiewicz2}
K. Bartkiewicz and A. Miranowicz, \pra \textbf{82}, 042330 (2010).

\bibitem{cloning1}
V. Scarani, S. Iblisdir, N. Gisin, and A. Acin, \rmp \textbf{77},
1225 (2005).

\bibitem{cloning2}
N.J. Cerf and J. Fiur\'a\v{s}ek, {\em Progress in Optics}, edited
by E. Wolf (Elsevier, Amsterdam, 2006), Vol. 49, pp. 455--545.


\bibitem{Adam2002}
A. Miranowicz, \c{S}. K. \"Ozdemir, Y.-X. Liu, M. Koashi, N.
Imoto, and Y. Hirayama, Phys. Rev. A {\bf 65}, 062321 (2002).

\bibitem{Sahin2002}
\c{S}.K. \"Ozdemir, A. Miranowicz, M. Koashi, and N. Imoto, \pra
\textbf{66}, 053809 (2002).

\bibitem{White1999}
A. G. White, D. F. V. James, P. H. Eberhard, and P. G. Kwiat, \prl
\textbf{83}, 3103 (1999).


\bibitem{Kwiat1995}
P. G. Kwiat, K. Mattle, H. Weinfurter, A. Zeilinger, and A. V.
Sergienko, and Y. Shih, \prl \textbf{75}, 4337 (1995).

\bibitem{Kwiat1999}
P. G. Kwiat, E. Waks, A. G. White, I. Appelbaum, and P. H.
Eberhard, \textbf{60}, R773 (1999).

\bibitem{Rangarajan2009}
R. Rangarajan, M. Goggin, and P. Kwiat, Opt. Express \textbf{17},
18920 (2009).

\bibitem{Tashima}
T. Tashima, S. K. \"Ozdemir, T. Yamamoto, M. Koashi, and N. Imoto,
\njp \textbf{11}, 023024 (2009).

\bibitem{Bohi}
P. B\"ohi, R. Prevedel, T. Jennewein, A. Stefanov, F. Tiefenbacher
and A. Zeilinger, Appl. Phys. B \textbf{89}, 499 (2007).

\bibitem{Prevedel2007}
R. Prevedel, P. Walther, F. Tiefenbacher, P. Böhi, R. Kaltenbaek,
T. Jennewein, and A. Zeilinger, Nature \textbf{445}, 65 (2007).


\bibitem{Pittman2001}
T.B. Pittman, B.C. Jacobs, and J.D. Franson, \pra \textbf{64},
062311 (2001).

\bibitem{Bartkowiak2010}
M. Bartkowiak and A. Miranowicz, \josab \textbf{27}, 2369 (2010).

\bibitem{Sahin2001}
\c{S}.K. \"Ozdemir, A. Miranowicz, M. Koashi, and N. Imoto, \pra
\textbf{64}, 063818 (2001); A. Miranowicz, J. Opt. B \textbf{7},
142 (2005).


\bibitem{Yamamoto2001}
T. Yamamoto, M. Koashi, and N. Imoto, \pra \textbf{64}, 012304
(2001).

\bibitem{Takeuchi1999}
S. Takeuchi, J. Kim, Y. Yamamoto, and H. H. Hogue, \apl
\textbf{74}, 1063 (1999).

\bibitem{Kaplan}
W. Kaplan, {\em Advanced Calculus} (Addison-Wesley, Reading, MA,
1992).

\bibitem{James2001}
D. F. V. James, P. G. Kwiat, W. J. Munro, and A. G. White, \pra
\textbf{64}, 052312 (2001).



\end{thebibliography}
\end{document}